# LHC at 10: THE PHYSICS LEGACY[1]

Michelangelo Mangano, Theoretical Physics Department, CERN, Geneva, Switzerland

Ten years have passed since the first high-energy proton-proton collisions took place at the Large Hadron Collider (LHC). Almost 20 more are foreseen for the completion of the full LHC programme. The data collected so far, from approximately 150 fb$^{-1}$ of integrated luminosity over two runs (Run 1 at a centre-of-mass energy of 7 and 8 TeV, and Run 2 at 13 TeV), represent a mere 5% of the anticipated 3000 fb$^{-1}$ that will eventually be recorded. But already their impact has been monumental.

Three major conclusions can be drawn from these first 10 years. First and foremost, Run 1 has shown that the Higgs boson — the previously missing, last ingredient of the Standard Model (SM) — exists. Secondly, the exploration of energy scales as high as several TeV has further consolidated the robustness of the SM, providing no compelling evidence for phenomena beyond the SM (BSM). Nevertheless, several discoveries of new phenomena *within* the SM have emerged, underscoring the power of the LHC to extend and deepen our understanding of the SM dynamics, and showing the unparalleled diversity of phenomena that the LHC can probe with unprecedented precision.

## EXCEEDING EXPECTATIONS

Last but not least, we note that 10 years of LHC operations, data taking and data interpretation, have overwhelmingly surpassed all of our most optimistic expectations. The accelerator has delivered a larger than expected luminosity, and the experiments have been able to operate at the top of their ideal performance and efficiency. Computing, in particular via the Worldwide LHC Computing Grid, has been another crucial driver of the LHC's success. Key ingredients of precision measurements, such as the determination of the LHC luminosity, or of detection efficiencies and of backgrounds using data-driven techniques beyond anyone's expectations, have been obtained thanks to novel and powerful techniques.

The LHC has also successfully provided a variety of beam and optics configurations, matching the needs of different experiments and supporting a broad research programme. In addition to the core high-energy goals of the ATLAS and CMS experiments, this has enabled new studies of flavour physics and of hadron spectroscopy, of forward-particle production and total hadronic cross sections. The operations with beams of heavy nuclei have reached a degree of virtuosity that made it possible to collide not only the anticipated lead beams, but also beams of xenon, as well as combined proton-lead, photon-lead and photon-photon collisions, opening the way to a new generation of studies of matter at high density.

Theoretical calculations have evolved in parallel to the experimental progress. Calculations that were deemed of impossible complexity before the start of the LHC have matured and become reality. Next-to-leading-order (NLO) theoretical predictions are routinely used by the experiments, thanks to a new generation of automatic tools. The next frontier, next-to-next-to-leading order (NNLO), has been attained for many important processes, reaching, in a few cases, the next-to-next-to-next-to-leading order (N$^3$LO), and more is coming [1].

Aside from having made these first 10 years an unconditional success, all these ingredients are the premise for confident extrapolations of the physics reach of the LHC programme to come [2].

To date, more than 2700 peer-reviewed physics papers have been published by the seven running LHC experiments (ALICE, ATLAS, CMS, LHCb, LHCf, MoEDAL and TOTEM). Approximately 10% of these are related to the Higgs boson, and 30% to searches for BSM phenomena. The remaining 1600 or so report measurements of SM particles and interactions, enriching our knowledge of the proton structure and of the dynamics of strong interactions, of electroweak (EW) interactions, of flavour properties, and more. In most cases, the variety, depth and precision of these measurements surpass those obtained by previous experiments using dedicated facilities. The multi-purpose nature of the LHC complex is unique, and encompasses scores of independent research directions. Here it is only possible to highlight a fraction of milestone results from the LHC's expedition so far.

---

[1] Published in *CERN Courier* March/April 2020 p 40, submitted to arXiv with permission.

# ENTERING THE HIGGS WORLD

The discovery by ATLAS and CMS of a new scalar boson in July 2012, just two years into LHC physics operations, was a crowning early success. Not only did it mark the end of a decades-long search, but it opened a new vista of exploration. At the time of the discovery, very little was known about the properties and interactions of the new boson. Eight years on, the picture has come into much sharper focus.

The structure of the Higgs-boson interactions revealed by the LHC experiments is still incomplete. Its couplings to the gauge bosons (W, Z, photon and gluons) and to the heavy third-generation fermions (bottom and top quarks, and tau leptons) have been detected, and the precision of these measurements is at best in the range of 5–10%. But the LHC findings so far have been key to establish that this new particle correctly embodies the main observational properties of the Higgs boson, as specified by the Brout-Englert-Guralnik-Hagen-Higgs-Kibble EW-symmetry breaking mechanism, referred hereafter as "BEH", a cornerstone of the SM. To start with, the measured couplings to the W and Z bosons reflect the Higgs' EW charges and are proportional to the W and Z masses, consistently with the properties of a scalar field breaking the SM EW symmetry. The mass dependence of the Higgs interactions with the SM fermions is confirmed by the recent ATLAS and CMS observations of the H→bb and H→ττ decays, and of the associated production of a Higgs boson together with a $t\bar{t}$ quark pair (see figure 1).

These measurements, which during the Run 2 of the LHC have surpassed the five-sigma confidence level, provide the second critical confirmation that the Higgs fulfills the role envisaged by the BEH mechanism. The Higgs couplings to the photon and the gluon (g), which the LHC experiments have probed via the H→γγ decay and the gg→H production, provide a third, subtler, test. These couplings arise from a combination of loop-level interactions with several SM particles, whose interplay could be modified by the presence of BSM particles, or interactions. The current agreement with data provides a strong validation of the SM scenario, while leaving open the possibility that small deviations could emerge from future higher statistics.

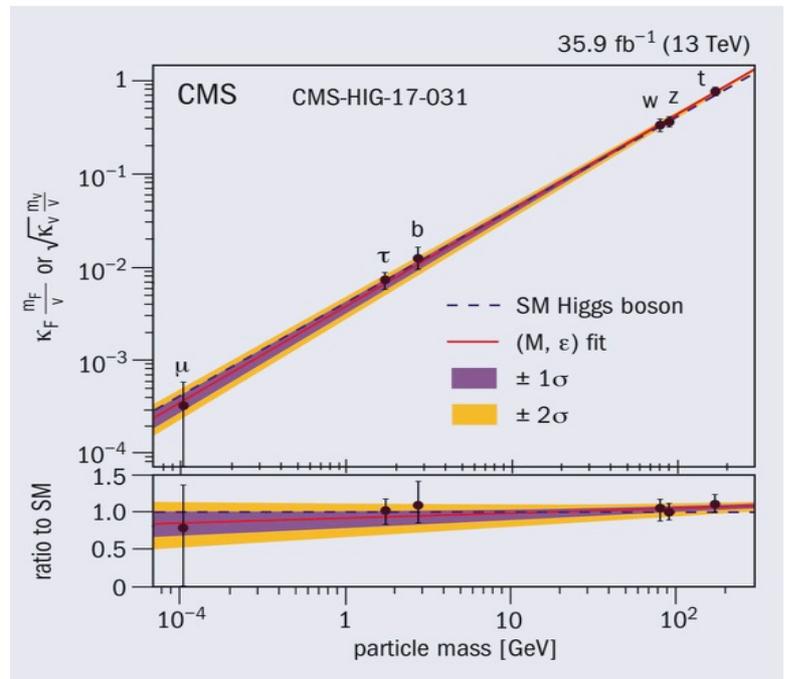

*Figure 1 Mass dependence of the Higgs-boson interactions with the SM fermions and massive gauge bosons, as confirmed by CMS and ATLAS (not shown) observations, revealing remarkable agreement with the predicted Yukawa interaction strength.*

The process of firmly establishing the identification of the particle discovered in 2012 with the Higgs boson goes hand-in-hand with two research directions, pioneered by the LHC: seeking the deep origin of the Higgs field, and using the Higgs boson as a probe of BSM phenomena. The breaking of the EW symmetry is a fact of nature, requiring the existence of a mechanism like BEH. But, if we aim beyond a merely anthropic justification for this mechanism (i.e. that, without it, physicists wouldn't be here to ask why), there is no reason to assume that nature chose its minimal implementation, namely the SM Higgs field. In other words: where does the Higgs boson detected at the LHC come from? This summarises many questions raised by the possibility that the Higgs boson is not just "put in by hand" in the SM, but emerges from a larger sector of new particles, whose dynamics induces the breaking of the EW symmetry. Is the Higgs elementary, or a composite state resulting from new confining forces? What generates its mass and self-interaction? More generally, is the existence of the Higgs boson related to other mysteries, such as the origin of dark matter (DM), of neutrino masses, or of flavour phenomena?

Ever since the Higgs-boson discovery, the LHC experiments have been searching for clues to address these questions, exploring a large number of observables [3]. All of the dominant production channels (gg fusion, associated production with vector bosons and with top quarks, and vector-boson fusion) have been discovered, and decay rates to WW, ZZ, γγ, bb and ττ were measured. A theoretical framework (effective field theory, EFT) has been developed to interpret in a global fashion all these measurements, setting strong constraints on possible deviations from the SM. With the larger data set accumulated during Run 2, the production properties of the Higgs have been studied with greater detail, simultaneously testing the accuracy of theoretical calculations, and the resilience of SM predictions.

To explore the nature of the Higgs boson, what has not been seen as yet can be as important as what was seen. For example, lack of evidence for Higgs decays to the fermions of the first and second generation is consistent with the SM prediction that these should be very rare. The H→μμ decay rate is expected to be about $3 \times 10^{-3}$ times smaller than that of H→ττ; the current sensitivity is two times below, and ATLAS and CMS hope to first observe this decay during the forthcoming Run 3, testing for the first time the Higgs coupling to the second-generation fermions. The SM Higgs boson is expected to conserve flavour, making decays such as H→μτ, H→eτ or t→Hc too small to be seen. Their observation would be a major revolution in physics, but no evidence has shown up in the data so far. Decays of the Higgs to invisible particles could be a signal of DM candidates, and constraints set by the LHC experiments are complementary to those from standard DM searches. Several BSM theories predict the existence of heavy particles decaying to a Higgs boson. For example, heavy top partners, T, could decay as T→Ht, and heavy bosons X decay as X→HV (V=W, Z). Heavy scalar partners of the Higgs, such as charged Higgs states, are expected in theories such as supersymmetry. Extensive and thorough searches of all these phenomena have been carried out, setting strong constraints on SM extensions.

As the programme of characterizing the Higgs properties continues, with new challenging goals such as the measurement of the Higgs self-coupling through the observation of Higgs pair production, the Higgs boson is becoming an increasingly powerful exploratory tool, to probe the origin of the Higgs itself, as well as a variety of solutions to other mysteries of particle physics.

**INTERACTIONS WEAK AND STRONG**

The vast majority of LHC processes are controlled by strong interactions, described by the quantum chromodynamics (QCD) sector of the SM. The predictions of production rates for particles like the Higgs or gauge bosons, top quarks, or BSM states, rely on our understanding of the proton structure, in particular of the energy distribution of its quark and gluon components (the parton distribution functions, PDFs). The evolution of the final states, the internal structure of the jets emerging from quark and gluons, the kinematical correlations between different objects, are all governed by QCD. LHC measurements have been critical, not only to consolidate our understanding of QCD in all its dynamical domains, but also to improve the precision of the theoretical interpretation of data, and to increase the sensitivity to new phenomena and to the production of BSM particles.

*Collisions galore*

Approximately $10^9$ proton—proton (pp) collisions take place each second inside the LHC detectors. Most of them bear no obvious direct interest for the search of BSM phenomena, but even the simple elastic collisions, pp→pp, which account for about 30% of this rate, have so far failed to be fully understood with first-principle QCD calculations. The ATLAS ALFA spectrometer and the TOTEM detector have studied these high-rate processes, measuring the total and elastic pp cross sections, at the various beam energies provided by the LHC. The energy dependence of the relation between real and imaginary part of the pp forward scattering amplitude has revealed new features, possibly described by the exchange of the so-called odderon, a coherent state of three gluons predicted in the 1970s [4].

The structure of the final states in generic pp collisions, aside from defining the large background of particles that are superimposed on the rarer LHC processes, is of potential interest to understand cosmic-ray (CR) interactions in the atmosphere. The LHCf detector measured the forward production of the most energetic particles from the collision, those driving the development of the CR air showers. These data are a unique benchmark to tune the CR event generators, reducing the systematics in the determination of the nature of the highest-energy CR constituents (protons, or heavy nuclei?), a step towards solving the puzzle of their origin.

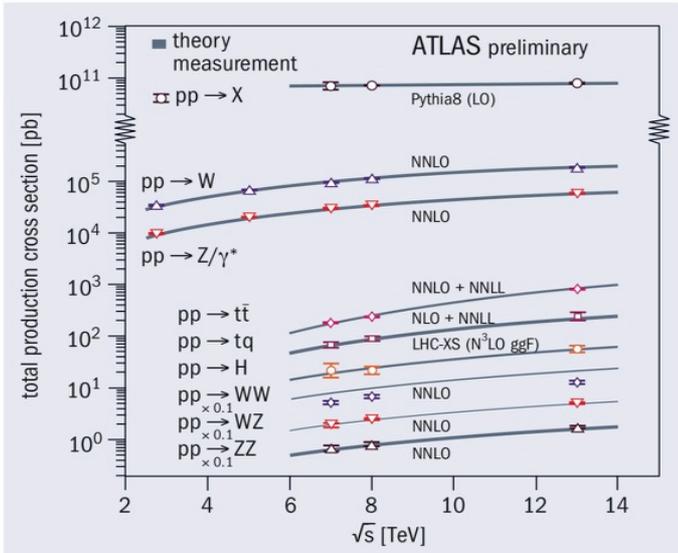 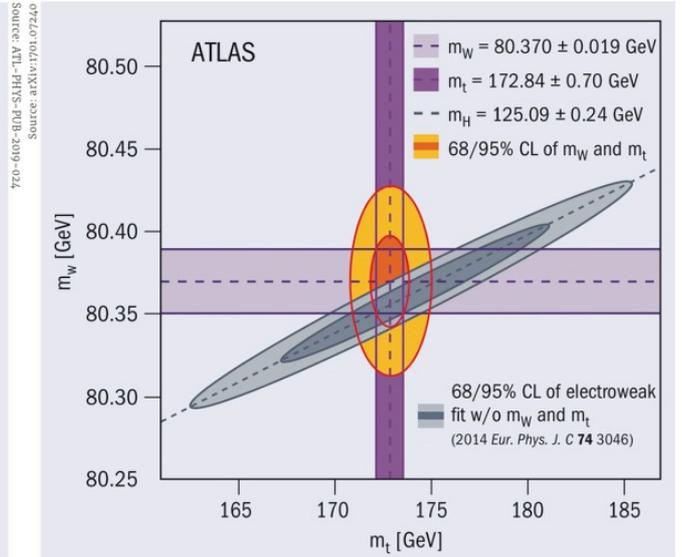

*Figure 2 Cross sections for key SM processes measured at different centre-of-mass energies, showing excellent agreement with state-of-the-art calculations.*

*Figure 3 ATLAS measurements of the W-boson and top-quark masses (horizontal and vertical bands, combined as orange contours) compared to their indirect determination from a global EW fit using the observed Higgs mass as input (grey contours).*

On the opposite end of the spectrum, rare QCD events with dijet pairs of mass up to 9 TeV have been observed by ATLAS and CMS. The study of their angular distribution, a Rutherford-like scattering experiment, has confirmed the point-like nature of quarks, down to $10^{-18}$ cm. The overall set of production studies, including gauge bosons, jets and top quarks, underpins countless analyses. Huge samples of top quark pairs, produced at 15 Hz, enable the surgical scrutiny of this mysteriously heavy quark, through its production and decays. New reactions, unobservable before the LHC, were first detected. Gauge boson scattering (e.g. $W^+ W^+ \rightarrow W^+ W^+$), a key probe of EWSB proposed in the 70s, is just one example. By and large, all data show an extraordinary agreement with theoretical predictions resulting from decades of innovative work (figure 2). Global fits to these data refine the proton PDFs, improving the predictions for the production of Higgs bosons or BSM particles.

The cross sections $\sigma$ of W and Z bosons provide the most precise QCD measurements, reaching a 2% systematic uncertainty, dominated by the luminosity uncertainty. Ratios like $\sigma(W^+)/\sigma(W^-)$ or $\sigma(W)/\sigma(Z)$, and the shapes of differential distributions, are known to a few parts in 1000. These data challenge the theoretical calculations' accuracy, and require caution to assess whether small discrepancies are due to PDF effects, new physics, or yet imprecise QCD calculations.

As already mentioned, the success of the LHC owes a lot to its variety of beam and experimental conditions. In this context, the data at the different centre-of-mass energies provided in the two runs are a huge bonus, since the theoretical prediction for the energy-dependence of rates can be used to improve the PDF extraction, or to assess possible BSM interpretations. The LHCb data, furthermore, cover a forward kinematical region complementary to that of ATLAS and CMS, adding precious information.

The precise determination of the W and Z production and decay kinematics have also allowed new measurements of fundamental parameters of the weak interaction: the W mass ($m_W$) and the weak mixing angle ($\sin\theta_W$). The measurement of $\sin\theta_W$ is now approaching the precision inherited from the LEP experiments and SLD, and will soon improve to shed light on the outstanding discrepancy between those two measurements. The $m_W$ precision obtained by the ATLAS experiment, $\Delta m_W$=19 MeV, is the best worldwide, and further improvements are certain. The combination with the ATLAS and CMS measurements of the Higgs boson mass ($\Delta m_H \cong 200$ MeV) and of the top quark mass ($\Delta m_{top} \lesssim 500$ MeV), provides a strong validation of the SM predictions (see figure 3). For both $m_W$ and $\sin\theta_W$ the limiting source of systematics is the knowledge of the PDFs, which future data will improve, underscoring the profound interplay among the different components of the LHC programme.

## QCD matters

The understanding of the forms and phases that QCD matter can acquire is a fascinating, broad and theoretically challenging research topic, which has witnessed great progress in recent years. Exotic multi-quark bound states, beyond the usual mesons ($q\bar{q}$) and baryons ($qqq$), were initially discovered at $e^+e^-$ colliders. The LHCb experiment, with its large rates of identified charm and bottom final states, is at the forefront of these studies, notably with the first discovery of heavy pentaquarks ($qqqc\bar{c}$) and with discoveries of tetraquark candidates in the charm sector ($qc\bar{c}\bar{q}$), accompanied by determinations of their quantum numbers and properties. These findings have opened a new playground for theoretical research, stimulating work in lattice QCD, and forcing a rethinking of established lore [5].

The study of QCD matter at high density is the core task of the heavy-ion programme. While initially tailored to the ALICE experiment, all active LHC experiments have since joined the effort. The creation of a quark–gluon plasma (QGP) led to astonishing visual evidence for jet quenching, with 1TeV jets shattered into fragments as they struggle their way out of the dense QGP volume. The thermodynamics and fluctuations of the QGP have been probed in multiple ways, indicating that the QGP behaves as an almost perfect fluid, the least viscous fluid known in nature. The ability to explore the plasma interactions of charm and bottom quarks is a unique asset of the LHC, thanks to the large production rates, which unveiled new phenomena such as the recombination of charm quarks, and the sequential melting of $b\bar{b}$ bound states.

While several of the qualitative features of high-density QCD were anticipated, the quantitative accuracy, multitude and range of the LHC measurements have no match [6]. Examples include ALICE's precise determination of dynamical parameters like the QGP shear-viscosity-to-entropy-density ratio, or the higher harmonics of particles' azimuthal correlations. A revolution ensued in the sophistication of the required theoretical modeling. Unexpected surprises were also discovered, particularly in the comparison of high-density states in PbPb collisions with those occasionally generated by smaller systems such as pp and pPb. The presence in the latter of long-range correlations, various collective phenomena and an increased strange baryon abundance (figure 4), resemble behavior typical of the QGP. Their deep origin is a mysterious property of QCD, still lacking an explanation. The number of new challenging questions raised by the LHC data is almost as large as the number of new answers obtained!

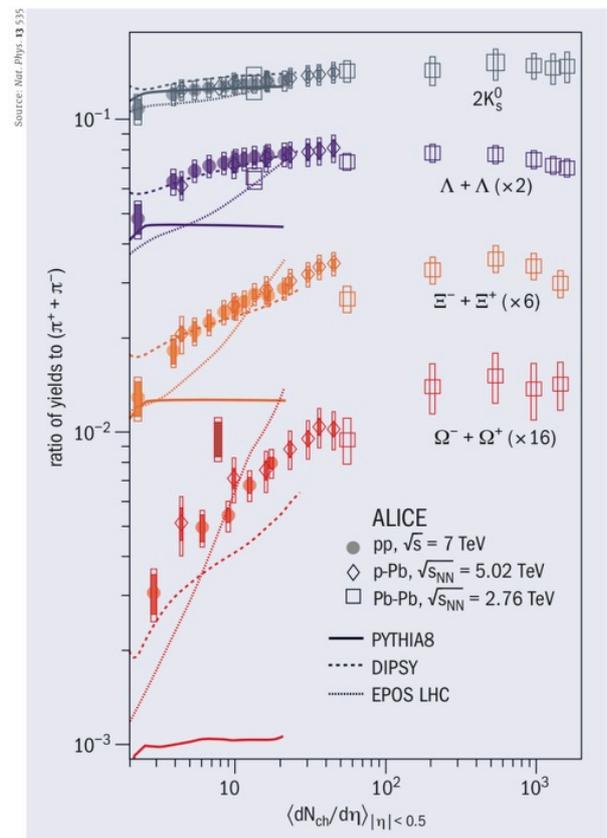

*Figure 4 Production yields for hadrons containing one (K,Λ), two (Ξ) and three (Ω) strange quarks, relative to the pion yield, as a function of the multiplicity of charged particles measured by ALICE in pp, pPb and PbPb collisions. The unexpected continuity across colliding systems is suggestive of the onset of a new class of collective phenomena for pp and pPb, progressively leading towards the PbPb behavior, which is attributed to the formation of the quark-gluon plasma.*

## FLAVOUR PHYSICS

Understanding the structure and the origin of flavor phenomena in the quark sector is one of the big open challenges of particle physics [7]. The search for new sources of CP violation, beyond those present in the CKM mixing matrix, underlies the efforts to explain the baryon asymmetry of the universe. In addition to flavour studies with Higgs bosons and top quarks, over $10^{14}$ charm and bottom quarks have been produced so far by the LHC, and the recorded subset has led to landmark discoveries and measurements. The rare $B_s \rightarrow \mu\mu$ decay, with a minuscule rate of approximately $3 \times 10^{-9}$, has been discovered by the LHCb, CMS and ATLAS experiments. The rarer $B_d \rightarrow \mu\mu$ decay is still unobserved, but its expected ~$10^{-10}$ rate is within reach. These two results alone had a big impact on constraining the parameter space of several BSM theories, notably supersymmetry, and their precision and BSM sensitivity will continue improving. LHCb has discovered $D\bar{D}$ mixing and the long-elusive CP violation in D-meson decays, a first for up-type quarks (figure 5). Large hadronic non-perturbative uncertainties make the interpretation of these results particularly challenging, leaving under debate whether the measured properties are consistent with the SM, or signal new physics. But the experimental findings are a textbook milestone in the worldwide flavour physics programme.

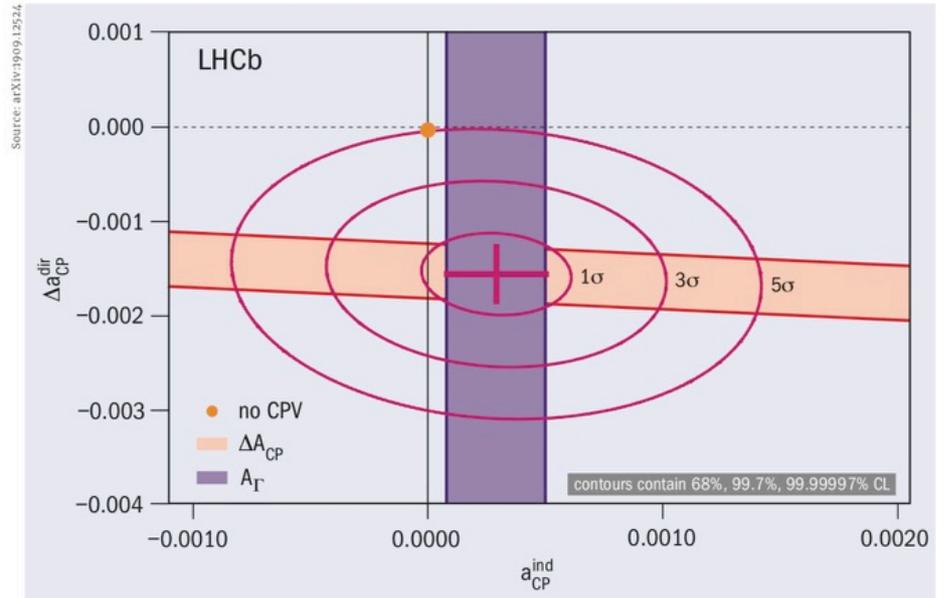

Figure 5 A combination of LHCb measurements of two types of CP violation in charm decays used to extract two underlying theory parameters. The vertical axis shows the difference in CP violation in two charm decays and here the combination (cross) is more than five standard deviations from the point of zero CP violation, constituting the first observation of CP violation in charm decays.

LHCb produced hundreds more measurements of heavy hadron properties and flavour-mixing parameters. Examples include the most precise measurement of the CKM angle $\gamma = (74.0^{+5.0}_{-5.8})°$ and, with ATLAS and CMS, the first measurement of $\phi_s$, the tiny CP-violation phase of $B_s \rightarrow J/\psi\phi$, whose precisely predicted SM value is very sensitive to new physics. With a few notable exceptions, all results confirm the CKM picture of flavour phenomena. Those exceptions, however, underscore the power of LHC data to expose new unexpected phenomena: $B \rightarrow D^{(*)}\ell\nu$ ($\ell=\mu,\tau$) and $B \rightarrow K^{(*)}\ell^+\ell^-$ ($\ell=e,\mu$) decays hint at possible deviations from the expected lepton flavour universality [8]. The community is eagerly waiting for further developments.

## BEYOND THE STANDARD MODEL

Years of model building, stimulated before and after the LHC start-up by the conceptual and experimental shortcomings of the SM (e.g. the hierarchy problem and the existence of DM), have generated scores of BSM scenarios to be tested by the LHC. Evidence has so far escaped hundreds of dedicated searches, setting limits on new particles up to several TeV (figure 6). Nevertheless, much was learned. While none of the proposed BSM scenarios can be conclusively ruled out, for many of them survival is only guaranteed at the cost of greater fine-tuning of the parameters, reducing their appeal. In turn, this led to rethinking the principles that implicitly guided model building. Simplicity, or the ability to explain at once several open problems, have lost some drive. The simplest realizations of BSM models relying on supersymmetry, for example, were candidates to at once solve the hierarchy problem, provide DM candidates and set the stage for the grand unification of all forces. If true, the LHC should have piled up evidence by now. Supersymmetry remains a preferred candidate to achieve that, but at the price of more Byzantine constructions. Solving the hierarchy problem remains the outstanding theoretical challenge. New ideas have come to the forefront, ranging from the Higgs potential being determined by the early-universe evolution of an axion field, to dark sectors connected to the SM via a Higgs portal. These latter scenarios could also provide DM candidates alternative to the weakly-interacting massive particles, which so far have eluded searches at the LHC and elsewhere.

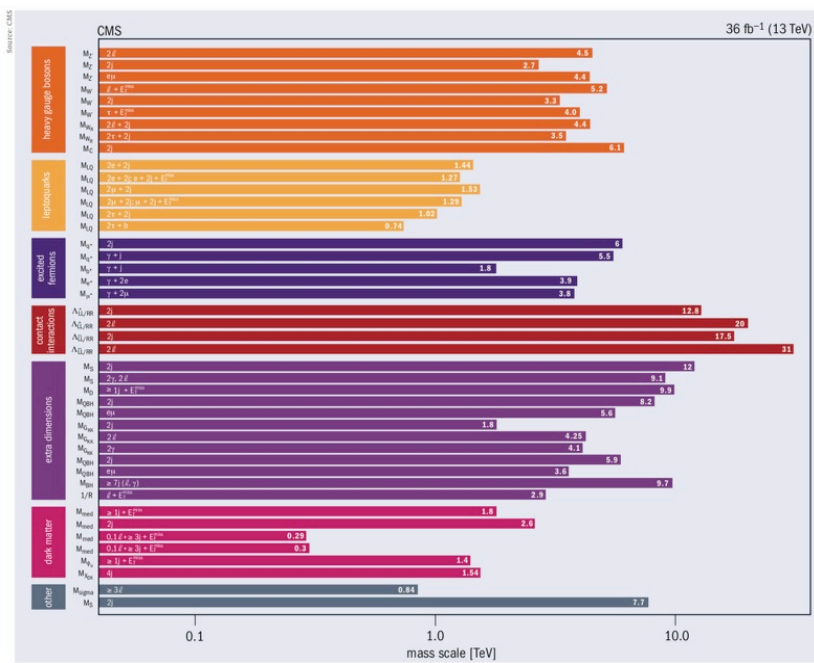

*Figure 6 Exclusion limits from CMS on the masses of certain exotic phenomena beyond the SM, using data collected in 2016, which extend to several TeV.*

With such rapid evolution of theoretical ideas taking place as the LHC data runs progressed, the experimental analyses underwent a major shift, relying on "simplified models": a novel model-independent way to represent the results of searches, allowing published results to be later reinterpreted in view of new BSM models. This amplified the impact of experimental searches, with a surge of phenomenological activity and the proliferation of new ideas. The cooperation and synergy between experiments and theorists have never been so intense.

Having explored the more obvious search channels, the LHC experiments refocused on more elusive signatures. Great efforts are now invested in searching corners of parameter space, extracting possible subtle signals from large backgrounds, thanks to data-driven techniques, and to the more reliable theoretical modelling that has emerged from new calculations and many SM measurements. The possible existence of new long-lived particles opened a new frontier of search techniques and of BSM models, triggering proposals for new dedicated detectors (Mathusla, CODEX-b and FASER, the last of which was recently approved for construction and operation in Run 3). Exotic BSM states, like the milli-charged particles present in some theories of dark sectors, could be revealed by MilliQan, a recently proposed detector. Highly ionizing particles, like the esoteric magnetic monopoles, have been searched for by the MoEDAL detector, which places plastic tracking films cleverly in the LHCb detector hall.

While new physics is still eluding the LHC, the immense progress of the last 10 years has changed forever our perspective on searches and on BSM model building.

**FINAL CONSIDERATIONS**

Most of the results only parenthetically cited, like the precision on the mass of the top quark, and others not even quoted, are the outcome of hundreds of years of person-power work, and would have certainly deserved more attention here. Their intrinsic value goes well beyond what was outlined, and they will remain long-lasting textbook material, until future work at the LHC and beyond improves them.

Theoretical progress has played a key role in the LHC's progress, enhancing the scope and reliability of the data interpretation. Further to the developments already mentioned, a deeper understanding of jet structure has spawned techniques to tag high-$p_T$ gauge and Higgs bosons, or top quarks, now indispensable in many BSM searches. Innovative machine-learning ideas have become powerful and ubiquitous. This article has concentrated only on what has already been achieved, but the LHC and its experiments have a long journey of exploration ahead.

The terms *precision* and *discovery*, applied to concrete results rather than projections, well characterize the LHC 10-year legacy. Precision is the keystone to consolidate our description of nature, increase the sensitivity to SM deviations, give credibility to discovery claims, and to constrain models when evaluating different microscopic origins of possible anomalies. The LHC has already fully succeeded in these goals. The LHC has also proven to be a discovery machine, and in a context broader than just Higgs and BSM phenomena. Altogether, it delivered results that could not have been obtained otherwise, immensely enriching our understanding of nature.